\documentclass[iop]{emulateapj}
\usepackage{epsfig}
\usepackage{apjfonts}
\usepackage{aas_macros}

\begin{document}

\title{Searching for magnetar powered merger-novae from short GRBs}
\author{He Gao$^{1}$, Bing Zhang$^{2,3,4}$, Hou-Jun L\"{u}$^{5,6}$ and Ye Li$^2$}
\affiliation{
$^1$Department of Astronomy, Beijing Normal University, Beijing 100875, China; gaohe@bnu.edu.cn\\
$^2$Department of Physics and Astronomy, University of Nevada Las Vegas, NV 89154, USA;\\
$^3$Department of Astronomy, School of Physics, Peking University, Beijing 100871, China; \\
$^4$Kavli Institute of Astronomy and Astrophysics, Peking University, Beijing 100871, China;\\
$^5$GXU-NAOC Center for Astrophysics and Space Sciences, Department of Physics, Guangxi University, Nanning 530004, China;\\
$^6$Guangxi Key Laboratory for Relativistic Astrophysics, Nanning, Guangxi 530004, China}

\begin{abstract}
The merger of a double neutron star (NS-NS) binary may result in a rapidly rotating massive NS with an extremely strong magnetic field (i.e., a millisecond magnetar). In this case, the magnetic spin-down of the NS remnant provides an additional source of sustained energy injection, which would continuously power the merger ejecta. The thermal emission from the merger ejecta would give rise to a bright optical ``magnetar-powered merger-nova". 
In this work, we carry out a complete search for magnetar-powered merger-nova from \emph{Swift} short gamma-ray burst (SGRB) sample. We focus on short GRBs with extended emission or internal plateau, which may signify the presence of magnetars as the central engine. We eventually find three candidates of magnetar-powered merger-nova from the late observations of GRB 050724, GRB 070714B and GRB 061006. With standard parameter values, the magnetar remnant scenario could well interpret the multi-band data of all three bursts, including the extended emission and their late chromatic features in the optical and X-ray data. The peak luminosities of these merger-novae reach several times $10^{42}~{\rm erg~s^{-1}}$, more than one order of magnitude brighter than the traditional ``kilo-novae'' with peak luminosity of $\sim 10^{41}~{\rm erg~s^{-1}}$. Intense, multi-color late time observations of short GRBs are encouraged to identify more merger-novae in the future.
\end{abstract}


%%%%%%%%%%%%%%%%%%%%%%%%%%%%%%%%%%%%%%%%
\section {Introduction}

The direct detection of gravitational waves (GWs) was recently achieved by the Laser Interferometer 
Gravitational-wave Observatory (LIGO) team, who reported two GW events (GW 150914 and GW 151226) and one more candidate (LVT 151012) from black hole (BH) binary mergers \citep{abbott16a,abbott16b}. A brand new observational window to study the physical universe has been opened. Because of the faint nature of GW signals, detecting an electromagnetic (EM) emission signal coincident with a GW signal in both trigger time and spatial direction is essential to confirm the astrophysical origin of the GW signals and study the astrophysical properties of the GW sources (e.g. host galaxy, distance). 

The frequency range of the ground-based GW detectors, such as Advanced LIGO \citep{ligo}, Advanced VIRGO \citep{virgo} and KAGRA \citep{kagra} interferometers, is designed to uncover the final inspiral and merger of compact object binary (NS-NS, NS-BH, BH-BH) systems. Whether a BH-BH merger system could have an EM counterpart still needs further observations (see a tentative candidate GW150914-GBM claimed by the {\em Fermi}/GBM team \citep{connaughton16} and counter opinions \citep{xiong16,greiner16}; and some theoretical models \citep{zhang16,loeb16,perna16} and counter opinions \citep{zhangsn16}). GW signals from NS-NS and NS-BH mergers, on the other hand, are highly expected to be associated with EM signals \cite[e.g.][for reviews]{metzger12,chu16}. The GW signals from NS-NS and NS-BH mergers may be detected by the current GW detectors (advanced LIGO and Virgo) in the near future.

The relative brightness of EM counterparts of a GW source is essentially determined by the properties of the leftover remnant from the merger. For NS-BH mergers, the merger remnant is a BH (surrounded by an accretion torus). NS-NS mergers, on the other hand, may result in four different types of merger remnants  \citep{rosswog00,dai06,fanxu06,gaofan06,rezzolla10,rezzolla11,zhang13,giacomazzo13,rosswog14,lasky14}: 1) a promptly formed BH; 2) a temporary, hyper-massive NS (supported by differential rotation) which survives 10-100 ms before collapsing into a BH; 3) a supra-massive NS supported by rigid rotation, which would survive to much longer time (e.g. longer than 100 s) before collapsing to a BH after the NS spins down; and 4) an indefinitely stable NS. The fractions for these outputs are essentially dependent on the total mass of the NS-NS system and the NS equation of state, which are currently not well constrained. Based on the constraints from  short Gamma-ray Burst (SGRB) data and one favored NS EoS, \cite{gao16} suggested that the fractions for the several outcomes of NS-NS mergers are as follows: $\sim40\%$ prompt BHs (or hypermassive NSs that collapse shortly after the mergers),  $\sim30\%$ supra-massive NSs that collapse to BHs in a range of delay time scales, and $\sim30\%$ stable NSs that never collapse. Overall, for NS-BH mergers and a fraction of NS-NS mergers ($\sim 40\%$), the post-merger product would be a BH surrounded by a mildly isotropic, sub-relativistic ejecta (which is composed of the tidally launched matter during the merger and the matter launched from the neutrino-driven wind from the accretion disk \cite[e.g.][]{rezzolla11,rosswog13,bauswein13,hotokezaka13}. The typical mass and speed
of the ejecta are in the range of $10^{-4}-10^{-2}~{\rm M}_{\odot}$ and $0.1-0.3~c$, respectively \citep{hotokezaka13}. In this case, the EM signals include a SGRB and its afterglow emission \citep{eichler89,narayan92,rosswog13,gehrels05,barthelmy05,berger11}, an optical/infrared transient  \citep{lipaczynski98,kulkarni05,metzger10,barnes13} powered by the radioactivity of the neutron-rich materials ejected during the coalescence process \citep{rezzolla11,rosswog13,hotokezaka13}, and a long-lasting radio emission as the ejecta interacts with the ambient medium \citep{nakar11,piran13}. The SGRB and its afterglow are relativistic and likely collimated \citep{burrows06,depasquale10} so that they are only detectable in preferred directions. On the other hand, the optical/infrared transient and the long-lasting radio emission are mildly isotropic and non-relativistic (due to heavy mass loading), and therefore may be detected from any direction if the flux is high enough. Since the optical/infrared transient is generated from the ejected materials during the merger process and is powered by radioactive decay from r-process radioactive materials, henceforth we call it an ``r-process-powered merger-nova". Note that such a transient is also named as ``macro-nova" by \cite{kulkarni05} due to its sub-supernova luminosity, or ``kilo-nova" by \cite{metzger10} due to its luminosity being roughly $\sim10^3$ times of the nova luminosity. 

Besides the black hole merger product, NS-NS mergers could also result in a supra-massive NS or even an indefinitely stable NS. In this case, magnetic spin-down of the NS remnant would provide an additional source of sustained energy injection, which greatly enriches and enhances the EM signals. A jet may be still launched via accretion from a disk to power the SGRB emission \citep{metzger08,dessart09,lee09,fernandez13,zhangd10}. The thermal emission from the ejecta would be significantly enhanced since heating by the magnetar wind could easily exceed the r-process power \citep{yu13,metzger14b,kasen15,siegel16a,siegel16b}, henceforth we call it a ``magnetar-powered merger-nova". Moreover, the magnetar-power would energize and accelerate the ejecta to a mildly or even moderately relativistic speed \citep{metzger14c}, giving rise to a strong broad-band afterglow-like emission upon interaction with the ambient medium (i.e. the double neutron star (DNS) merger afterglow model, \citealt{gao13GWB}). Finally, in the directions with ejecta cavity or after the ejecta becomes transparent, X-rays from direct dissipation of the magnetar wind could escape and reach the observer \citep{zhang13,sun16} . 

After several years of search from new observations and archival data, some of the proposed signals were claimed to be detected or to be associated with some short GRBs. For instance, a rapidly fading optical transient source, PTF11agg, was reported by the Palomar Transient Factory (PTF) team, which was proposed to be a good candidate for DNS merger afterglow emission \citep{wanglj13,wu14}. A r-process-powered merger-nova was claimed to be detected in the near-infrared band for GRB 130603B\footnote{Prompted by this detection,  \cite{yang15} re-examined the late afterglow data of GRB 060614 observed with HST, and claimed to find another candidate r-process-powered merger-nova associated with this event. It is worth noticing that according to its duration, GRB 060614 is classified as a long burst. However, its multi-wavelength observational properties are more consistent with those GRBs with a compact object merger origin \citep{gehrels06,gal-yam06,zhang07}. }  \citep{tanvir13,berger13,fan13a}. Meanwhile, \cite{gao15} made a comprehensive analysis on the multi-band observations of GRB 080503, and suggested a case to connect the late-time optical excess of GRB 080503 \citep{perley09} with a magnetar-powered merger-nova (see also an earlier qualitative proposal by \cite{metzger14}). Unfortunately this burst did not have a redshift measurement, so that the precise energetics of the merger-nova cannot be measured. Some rapidly evolving and luminous transients discovered recently with the Pan-STARRS1 Medium Deep Survey was proposed by \cite{yu15} as good candidates of magnetar-powered merger-nova. Most recently, a complete search for r-process-powered merger-nova had been made by comparing the late-time optical-nIR data of all short GRBs to r-process-powered merger-nova models \citep{fong16b,jin16}, and \cite{jin16} claimed to find one more candidate in GRB 050709. 
 
In this work, we attempt to carry out a complete search for magnetar-powered merger-novae. 

Evidence of a magnetar following some SGRBs has been collected in the Swift data. For instance, about 20\% of short GRBs detected with {\em Swift} have soft $\gamma$-ray extended emission \citep{norris06,sakamoto11} following their initial short, hard spike. Some other SGRBs show an internal X-ray plateau followed by very rapid decay \citep{rowlinson10,rowlinson13,lv15}, which requires a long-lasting central engine. A statistical analysis suggests that the so-called extended emission is consistent with being the brighter, harder internal X-ray plateau emission \citep{lv15}. All these features are difficult to interpret within the framework of a black hole central engine, but are consistent with the existence of a rapidly spinning millisecond magnetar central engine that survives for an extended period of time before collapsing\citep{rowlinson10,rowlinson13,gompertz13,gompertz14,lv14,lv15}. We thus focus on the SGRB sample with extended emission or X-ray internal plateau emission in order to search for magnetar-powered merger-novae\footnote{Systematical studies for magnetar-powered radio emission from similar SGRB samples have already been performed in various works \citep{metzger14c,horesh16,fong16}.}. In Section 2, we give a general description of different emission sites and emission components involved in the scenario when NS-NS mergers leave behind a NS.  In Section 3, we present our systematic search of merger-novae from the SGRB sample and show three candidates, GRB 050724, GRB 070714B and GRB 061006, whose multi-band observations can be well fitted by the magnetar model. The conclusion and implications of our results are discussed in Section 4. Throughout the paper, the convention $Q = 10^nQ_n$ is adopted in cgs units.

\section{Model description}

If the total mass of binary neutron star system is small enough and the equation of state of nuclear matter is stiff enough, the merger of two NSs could leave behind an indefinitely stable or a supra-massive NS \citep{dai06,gaofan06,zhang13,lasky14,fryer15,li16}. Considering that before the merger the two NSs are in the Keplerian orbits, the newborn massive NS should be rapidly spinning near the breakup limit and may also carry a strong magnetic field $B \gtrsim 10^{14}\,{\rm G}$ similar to that of known Galactic ``magnetars" \cite[][and reference therein]{zhang13,metzger14}. The millisecond magnetar is surrounded by a mildly isotropic, sub-/mildly-relativistic ejecta.  A quasi-spherical symmetry for the ejecta could be reasonably assumed considering a variety of origins of the ejecta materials \citep{metzger14}. Under this scenario, there are four emission sites and several emission components (see details in \cite{gao15}). 

\begin{itemize}
\item The magnetar would be initially surrounded by a centrifugally supported accretion disc
\citep{metzger08,dessart09,lee09,fernandez13}, so that a short-lived ($\lesssim~{\rm s}$),
collimated jet could be launched \citep{zhangd10,bucciantini12,nagakura14}. The jet component powers the short spike in prompt emission and the GRB afterglow emission.
\item After the jet breaks out from the ejecta, the hole punched by the jet remains open as the Poynting-flux-dominated magnetar wind continuously penetrates through the hole. An early magnetar wind component powers the soft extended emission and the high latitude tail emission.
\item In a wider region, the magnetar wind encounters the ejecta, and a significant fraction of the wind energy (parameterized as $\xi$) would be deposited into the ejecta. Such a continuous energy injection process not only heats the ejecta material to power the merger-nova \citep{yu13,metzger14,metzger14c}, but also accelerates the ejecta to a higher speed (even reaching a mildly or moderately relativistic speed in some parameter regimes), giving rise to bright afterglow emission by driving a strong forward shock into the ambient medium \citep{gao13GWB}. 
\item After the characteristic spindown time scale, the magnetic pressure drops quickly, so that the hole punched by the initial jet would be closed. The remaining fraction of the wind energy ($1-\xi$) would be deposited into the ejecta in the form of thermal and kinetic energy.  The dissipation photons in the magnetar wind or in the wind-envelope interaction regions would be trapped but would eventually diffuse out with a reducing factor $e^{-\tau}$, where $\tau$ is the optical depth of the ejecta. At a later epoch, the entire ejecta becomes optically thin. The X-rays powered by magnetar-wind dissipation would escape freely and reach the observer directly.
\end{itemize}

A detailed description of the model is presented in the Appendix.

%%%%%%%%%%%%%%%%%%%%%%%%%%%%%%%%%%%%%%%%
\section{Candidate search}
\subsection{Sample selection}

For a complete search of magnetar-powered merger-novae, we set up three criteria for sample selection:
\begin{itemize}
\item We focus on short GRBs with extended emission or internal plateau, which may signify the presence of a supra-massive magnetar at the central engine.
\item We focus on bursts with high-quality late-time data in both X-ray and optical bands. More specifically, at time $\geq 10^4$ s after the trigger, a candidate should have enough photons to extract a reasonable X-ray light curve and it should have at least two detections (and/or with several upper limits) in the optical band. We require that the late optical data and X-ray data should clearly show a feature deviating from the standard external shock model, for instance, a re-brightening feature showing up in the optical lightcurve and/or in the X-ray lightcurve, or the power-law indices for these two bands are clearly deviating from the so-called closure-relations, suggesting that the X-ray and optical data are contributed by different emission components.
\item In order to obtain quantitatively fitting of the data, we require the bursts to have redshift measurements. 
\end{itemize}
Based on these criteria, we systematically investigate 96 {\em Swift} short GRBs from the launch of {\em Swift} to October 2015. We first search for those bursts with extended emission or an internal plateau (which shares the same physical origin as shown by \cite{lv15}). The data analysis details have been presented in several previous papers \citep{zhangbb07,liang07,lv15}. Basically, by assuming a single power-law spectrum \citep{obrien06,willingale07,evans09}, we extrapolate the BAT data to the XRT band and then perform a temporal fit to the XRT-band light curve with a smooth broken power law in the rest frame to identify a possible plateau (defined as a temporal segment with decay slope shallower than 0.5). We find 21 bursts that exhibit a plateau followed by a decay slope steeper than 3 as our ``internal plateau'' sample. Within this sample, 5 bursts have high quality detection data in both X-ray and optical bands at epochs later than $10^4$ s after the prompt emission trigger. These are GRB 050724, GRB 051227, GRB 061006, GRB 070714B, and GRB 080503. The case of GRB 080503 has been separately discussed in \cite{gao15}. We exclude GRB 051227 from the following analysis since there is no redshift detection for this burst. Interestingly, the remaining three bursts all exhibit chromatic behaviors in late optical and X-ray observations. In the following, we discuss these three bursts in details.

\subsection{Case study}
\subsubsection{GRB\,050724}

GRB\,050724 was detected by Burst Alert Telescope (BAT) on abroad \emph{Swift} at 12:34:09 UT
on 2005 July 24. The prompt BAT light curve contains a short-lived spike, with an FWHM of $\sim 0.25$ s \citep{GCN.3665,GCN.3667}, which is followed by soft extended emission lasting for at least 200 s \citep{barthelmy05,campana06}. The total fluence was $(6.3\pm1.0)\times 10^{-7}~\rm ergs~cm^{-2}$ \citep{GCN.3667} over the $15-350$ KeV band. \cite{GCN.3679} identified the host galaxy of the transient to be a massive early-type galaxy at a redshift of $z=0.258\pm0.002$.  Spectroscopic redshifts were reported by \cite{GCN.3700} ($z=0.258\pm0.002$) and \cite{berger05} ($z=0.257\pm0.001$).

Following the early very steep decay between 100 and 300 s \citep{campana06}, the X-ray light curve
is well fit by a single underlying power-law decay slope of $\alpha = 0.98$, where $F_{\nu}
\propto t^{-\alpha}$, then the X-ray flux rebrightened to the level of $3\times 10^{-12}~ {\rm erg~s^{-1}~cm^{-2}}$ around $4\times 10^{4}~{\rm s}$ after the BAT trigger \citep{margutti11}. 

In the optical band, the Russian-Turkish 1.5-m telescope observed the field of GRB 050724 around $2\times10^4$ s after the prompt trigger, and provides an upper limit of $\lesssim 4~\rm{\mu Jy}$ in R band \citep{GCN.3671}. Then, the field was observed with the ESO Very Large Telescope (VLT), using the FORS1 instrument, starting 0.5 days after the GRB. The afterglow is detected up to 3.5 days after the GRB. For the VLT observations only, assuming a power-law behavior, the decay slope in R band is $\alpha_R=1.51\pm0.09$ \citep{malesani07}. Considering the 1.5-m telescope upper limit information and the VLT detections, the optical afterglow of GRB 050724 showed a re-brightening feature from $\sim10^4$ s to $\sim10^5$ s rather than simple power-law decay. 

We investigate the broadband data of GRB 050724 with the physical model described in the section 2. The soft extended emission and the late X-ray re-brightening can be connected with a magnetar spin-down luminosity evolution function, suggesting direct magnetic dissipation as the same underlying origin for these two observed components. In the X-ray band, after the extended emission and the early steep decay, the funnel (i.e., the hole punched by the initial jet) is closed due to the sideway pressure \citep{gao15} so that one is left with the X-ray emission from the external shock. The afterglow decay slope during this period of time is about $\alpha = 0.98$. In the optical band, the re-brightening feature suggesting that the optical data may not be contributed by the GRB external-shock afterglow component. Below we show that it can be explained by the emission from a magnetar-powered merger-nova (see Figure 1 for details). 

Considering the relatively large parameter space and the fact that model parameters obtained by fitting GRB afterglow data usually suffer severe degeneracy \citep{kumarzhang15}, for the purpose of this work, we do not attempt to fit the data across a large parameter space. For each candidate, we present a set of parameter values that could interpret the data well. For GRB 050724, the jet isotropic kinetic energy $E_k$ is set to $3.9\times10^{50}~{\rm erg}$ and the ambient medium density $n$ is set to $0.1~{\rm cm^{-3}}$. The values for initial Lorentz factor ($\Gamma_0$) and half opening angle ($\theta$) of the jet are chosen as 200 and 0.2, respectively, and our final fitting results are not sensitive to these values. For microphysics shock parameters (the electron and magnetic energy fraction parameters $\epsilon_e$ and $\epsilon_B$, and the electron spectral index $p$), we choose their commonly used values in GRB afterglow modeling, i.e., $\epsilon_e=0.025$, $\epsilon_B=0.001$, and  $p=2.3$ \cite[][for a review]{kumarzhang15}\footnote{ The distribution of the $\epsilon_B$ value is wide, but growing evidence from afterglow modeling suggests that many GRB afterglows have $\epsilon_B$ as low as $<10^{-5}$ \citep{santana14,wang15,beniamini15}.}. For the magnetar, the stellar radius $R_{\rm s}$ is adopted as $1.2\times 10^6$ cm. The initial spin period $P_i$ is taken as $5~{\rm ms}$ by
considering angular momentum loss via strong gravitational radiation \citep{fan13b,gao16,lasky16,radice16}. The dipolar magnetic field of strength $B$ is adopt as $6\times 10^{15}~{\rm G}$, which is consistent with the suggested values by fitting the SGRBs X-ray plateau feature \citep{rowlinson13,lv14}. For the ejecta, we take the standard values of mass ($M_{\rm ej}\sim10^{-3}{\rm M_{\odot}}$) and initial velocity ($v_i=0.2c$), and a mild value of the effective opacity $\kappa=1~{\rm cm^{2}~g^{-1}}$.  Finally, we assume a relatively small value ($1\%$) of the wind energy that is deposited into the ejecta, as suggested by \cite{metzger14}. See a collection of parameters in Table 1. One can see that the model can well interpret both the X-ray and optical lightcurves of the event.

\begin{figure*}[t]
\begin{center}
\begin{tabular}{lll}
\resizebox{60mm}{!}{\includegraphics[]{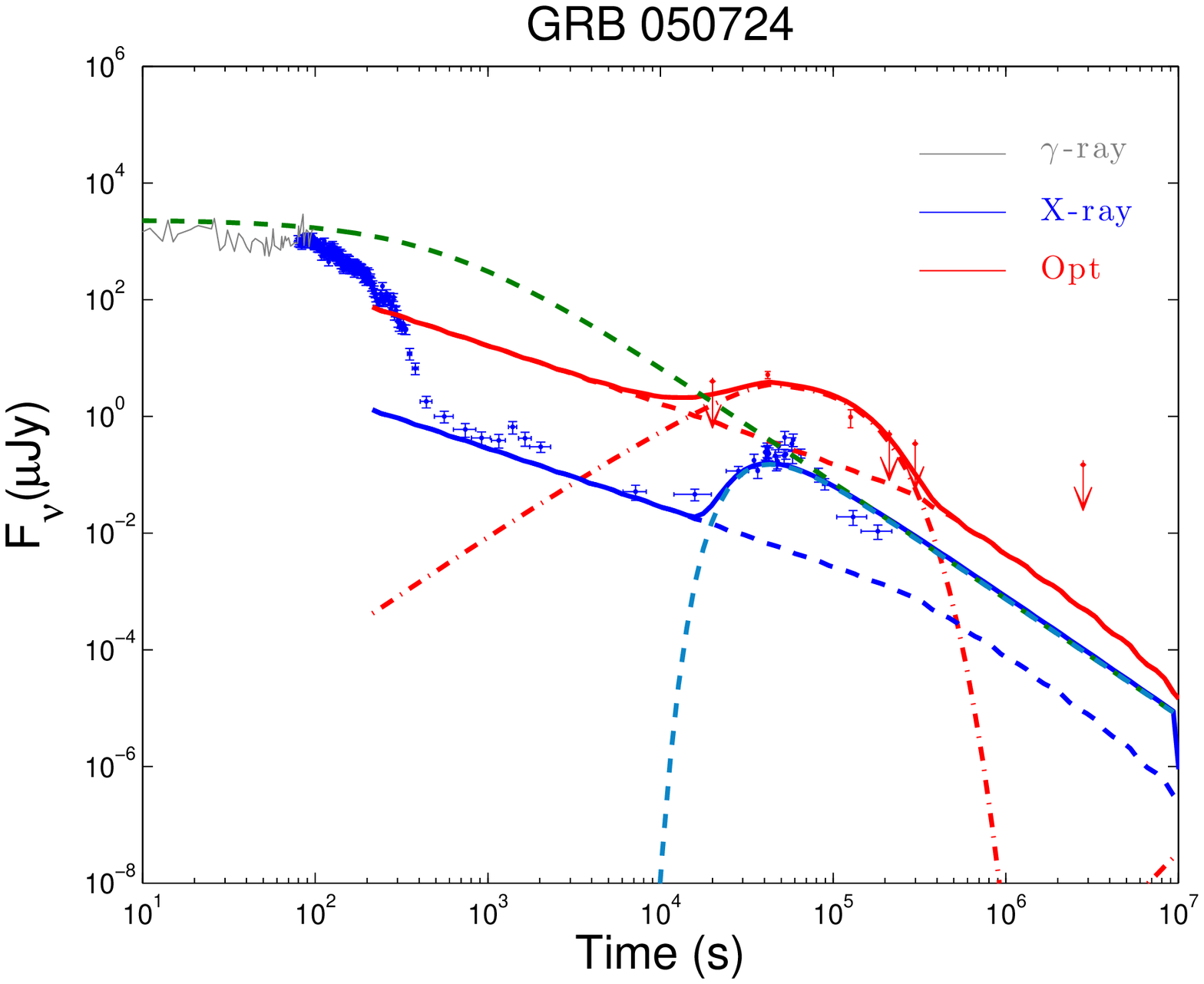}} &
\resizebox{60mm}{!}{\includegraphics[]{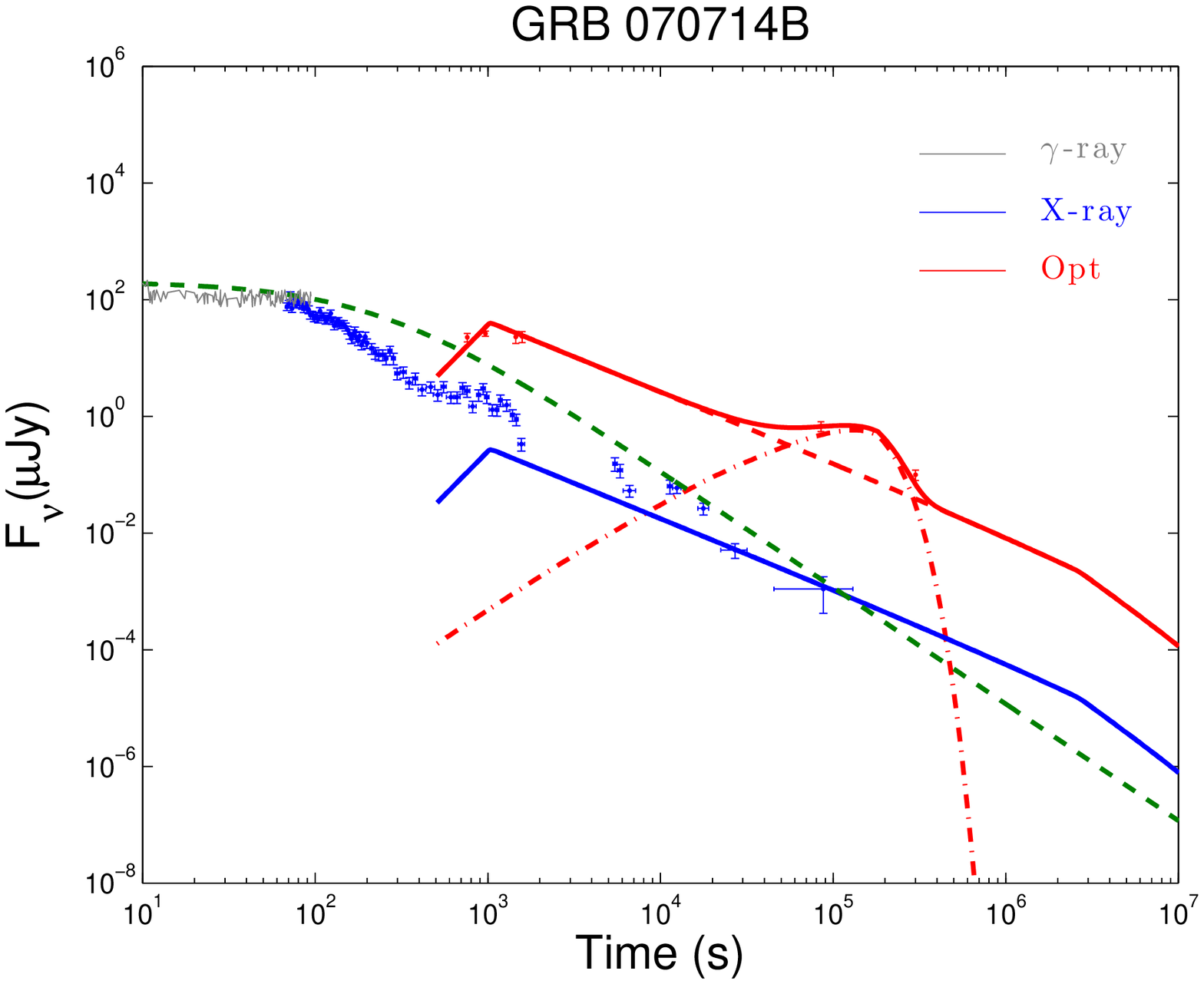}} &
\resizebox{60mm}{!}{\includegraphics[]{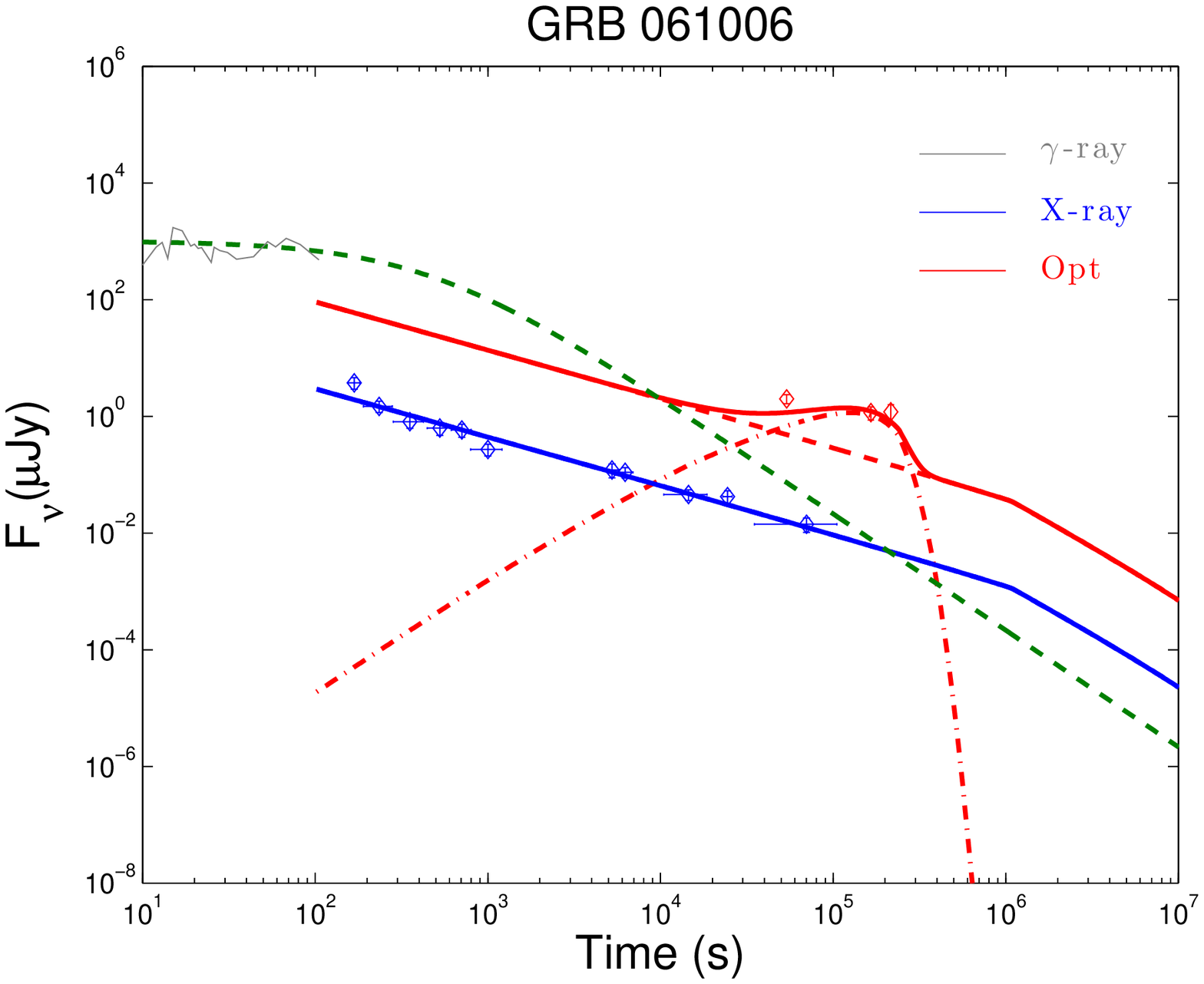}} \\
\end{tabular}
\caption{Modeling results for the broad-band observations of GRB 050724 (left panel), GRB 070714B (middle panel) and GRB 061006 (right panel). The time is in the observer frame. The blue and red colors denote X-ray and optical, respectively. Detections are denoted as dots or diamonds with error bars, and upper limits are denoted by downwards arrows. The dashed lines represent the GRB afterglow emission and the dotted dash lines represent the merger-nova emission. The evolution function of the magnetar spin-down radiation luminosity is presented by the green dashed line and the late magnetar wind dissipation emission is presented by the light blue dashed line. The solid lines denote the sum of various emission components. The optical points are not corrected for Galactic or intrinsic extinction. The upper limit data for GRB 050724 are at the 3$\sigma$ confidence level \citep{malesani07}.}
\label{fig:fit}
\end{center}
\end{figure*}

\subsubsection{GRB\,070714B}

GRB\,070714B was detected by the {\em Swift}-BAT at 04:59:29 UT on 14 July 2007 \citep{GCN.6620}.  The 15-150\,keV fluence in the prompt emission was measured to be $(5.1 \pm 0.3) \times 10^{-7}$\,erg cm$^{-2}$.  The prompt emission light curve is comprised of several short spikes (duration $\sim 3$\,s) with a long, soft tail 
\citep{GCN.6623}. The total duration is significantly longer than the canonical long-short divide ($T_{90} = 64 \pm 5$\,s).  The spectral lag is consistent with zero. GRB\,070714B therefore appears likely to be a member of the short-duration bursts with extended emission (e.g.~\cite{barthelmy05}). A fading optical afterglow was discovered inside the XRT error circle \citep{GCN.6627} by several groups \cite[][reference therein]{cenko08}.  \cite{cenko08} reported a single bright emission line at $\lambda = 7166.2 \pm 0.4$, which was identified as [\ion{O}{2}] $\lambda$ 3727 at $z = 0.9225 \pm 0.0001$.  This confirms the spectroscopic redshift first proposed by \citet{GCN.6836}.  

The XRT light curve shows a fading behavior with super-imposed small flaring.  The light curve can be fit with a power-law starting with a steep decay with a slope of $\alpha = 2.49\pm0.18$ followed by a plateau starting from  $\sim 400$ s with a slope of  $\alpha = 0.60\pm0.29$, which breaks at $\sim 1000$ s to a steeper decay of  $\alpha = 1.73\pm0.11$ \citep{GCN.6627}. 

Early optical afterglow decay was roughly flat or decaying with $\alpha \sim 0.07 \pm 0.28$ (assuming a power-law decay of the form $F(t) \propto t^{-\alpha}$).  This can be compared to the X-ray decay which shows a plateau during this period (see Figure 1) . Including the late time WHT observation, the optical decay rate becomes $\alpha = 0.86 \pm 0.10$, which is much shallower than the X-ray decay over a similar time frame of $\alpha = 1.73 \pm 0.11$ \citep{GCN.6630,GCN.6627}. Keck I was employed to observe the field of GRB 070714B about $3\times 10^5$ s after the prompt trigger, providing an detection point of $\sim0.1~\rm{\mu Jy}$ in R band \citep{GCN.6652}. With this information, the WHT data could no longer be fitted with a simple power-law decay, at least one break feature is required before $3\times 10^5$ s. In any case, whereas the early optical data points are roughly flat during the X-ray plateau so that the two bands may share a common origin, the optical and X-ray behaviors clearly diverge at late times \citep{graham09}.

The multi-band data of GRB 070714B could also be well interpreted with the physical model described in the section 2. 
Similar to GRB 050724, the late optical data point should be dominated by the emission from a magnetar-powered merger-nova, which explains the diverse behavior between X-ray and optical data. The early optical data corresponds to the onset phase of external shock emission, which explains the very flat decay index. After the steep decay, the baseline of the decaying X-ray data could be fitted by the external shock emission, while the early soft extended emission and the small flaring features superposed on the external shock emission may be contributed by the direct magnetic dissipation of the magnetar wind. 

In the GRB 070714B modeing, following parameters are adopted (Table 1): jet isotropic kinetic energy $E_k = 1.0\times10^{52}~{\rm erg}$, ambient medium density $n = 0.01~{\rm cm^{-3}}$, initial Lorentz factor $\Gamma_0 = 95$, half opening angle $\theta= 0.2$, shock parameters $\epsilon_e= 0.06$, $\epsilon_B= 0.0002$, $p= 2.6$,  neutron star radius $R_{\rm s}=1.0\times 10^6$ cm, magnetar initial spin period $P_i=2.5~{\rm ms}$, dipolar magnetic field of strength $B = 1\times 10^{16}~{\rm G}$, ejecta mass $M_{\rm ej}\sim10^{-2}~{\rm M_{\odot}}$, ejecta initial velocity $v_i=0.2c$, effective opacity as $\kappa=1~{\rm cm^{2}~g^{-1}}$, and $10\%$ efficiency for the wind energy deposition into the ejecta. 

\subsection{GRB\,061006}

GRB\,061006 was detected by {\em Swift}-BAT at 16:45:50 UT
on 2006 October 6 \citep{schady06}. This burst began with an intense double-spike from T-22.8 to T-22.3 seconds. This spike was also seen as a short GRB by RHESSI, Konus, and Suzaku \citep{GCN.5702}.  This was followed by lower-level persistent emission, making the total prompt duration as $130 \pm 10$\,s. The fluence in the 15-150 keV band is $(1.43\pm0.14)\times 10^{-6}~\rm ergs~cm^{-2}$ \citep{schady06}. \cite{berger07} obtained spectroscopic observations of a putative host galaxy with GMOS on Gemini-South on 2006 November 20.31 UT for a total exposure time of 3600 s, and detected weak continuum emission and several emission lines corresponding to [O ii] $\lambda$3727, H$\beta$, [O iii]$\lambda$4959, and [O iii]$\lambda$5007 at $z = 0.4377 \pm 0.0002$. 

The X-ray lightcurve shows an initial slope of $\alpha \sim2.26\pm0.1$, breaking around 290 s after the burst to a flatter decay slope of $\alpha \sim0.77\pm0.07$. Optical afterglow was observed with the ESO-VLT UT2 equipped with FORS1 twice \citep{GCN.5705,GCN.5718}, on 2006 Oct 7.30728 UT (14.6 hr after the prompt trigger) and on 2006 Oct 8.2936 UT (1.60 days after the prompt trigger).  The inferred power-law decay slope is quite shallow ($\alpha \sim0.50\pm0.08$).

Similar to GRB 050724 and GRB 070714B, the diverse behavior between X-ray and optical band at late times again suggests that GRB 061006 optical lightcurve may be contaminated by the emission from a magnetar-powered merger-nova. After the steep decay, the simple decaying X-ray data could be fitted by the external shock emission, while the early soft extended emission is contributed by the direct magnetic dissipation of the magnetar wind. Unlike GRB 050724 that showed a re-brightening in X-rays, no X-ray bump is seen near the peak of the optical merger nova. This might be explained by assuming that the supra-massive neutron star already collapsed before $10^4$ s.

In the interpretation of GRB 061006, the following parameters are adopted (Table 1): jet isotropic kinetic energy $E_k=1.6\times10^{52}~{\rm erg}$, ambient medium density $n=0.1~{\rm cm^{-3}}$, initial Lorentz factor $\Gamma_0= 200$, half opening angle $\theta= 0.2$, $\epsilon_e= 0.015$, $\epsilon_B= 0.00003$, $p= 2.1$,  neutron star radius $R_{\rm s}=1.0\times 10^6$ cm, initial spin period $P_i=2~{\rm ms}$, dipolar magnetic field of strength $B=5\times 10^{15}~{\rm G}$, ejecta mass $M_{\rm ej}\sim 10^{-2}{\rm M_{\odot}}$, ejecta initial velocity $v_i=0.2c$, effective opacity $\kappa=1~{\rm cm^{2}~g^{-1}}$, and $1\%$ wind deposition efficiency. 

\subsection{Summary}

The late time optical data of all three GRBs show a common feature of re-brightening, which can be well interpreted as the presence of a magnetar-powered merger-nova in each of them. The X-ray behaviors of the three bursts, on the other hand, are completely different. For instance, unlike GRB 050724, GRB061006 did not show a late re-brightening feature in the X-ray band, suggesting that the magnetar may have collapsed into a black hole before the surrounding ejecta becomes transparent\footnote{In the interpretation of GRB 070714B and GRB 061006, we invoke the magnetar collapsing time as an additonal free parameter $t_{\rm col}$. We find that the adopted value of  $t_{\rm col}$ barely affects the final results as long as it is much larger than the spin down timescale of the magnetar but smaller than the transparent timescale of the ejecta.}. On the other hand, the XRT light curve of GRB 070714 shows small flaring features superposed on the fading power-law behavior, which dose not exist for GRB 050724 and GRB 061006. The flaring is consistent with the erratic activity of a magnetar. Because of these repeated activities, the funnel punched by the jet never completely closes in contrast to GRB 050724 and GRB 080503. This can also explain the lack of a very steep decay phase in this burst, in contrast to the other two GRBs. In any case, all three cases can be interpreted within the framework of the magnetar-powered merger-nova model. 
 
It is worth noting that the late optical data points for GRB 061006 are close to each other in log space, and the last data is possibly contaminated by the host galaxy \citep{GCN.5718}. Taking into account that the X-ray light curve of GRB 061006 behaves as a simple power-law decay without any additional features after the initial decay phase, we put GRB 061006 as a less robust case compared with GRB 050724 and GRB 070714B.

%%%%%%%%%%%%%%%%%%%%%%%%%%%%%%%%%%%%%%%%

\section{Conclusions and Discussion}
\label{sec:discussion}

In this work, we carry out a complete search for magnetar-powered merger-nova from the short GRB data. 
With the three criteria we set up (extended emission / internal plateau, late time high-quality X-ray and optical data, and redshift), we are left with three bursts i.e., GRB 050724, GRB 070714B, GRB 061006. Interestingly, all three bursts exhibit chromatic behaviors in late optical and X-ray observations, suggesting that the X-ray and optical data are contributed by different emission components. In particular, the late optical data of the three bursts all show a clear bump, which is consistent with the presence of a merger-nova. The X-ray data of the three bursts show different behaviors (GRB 050724 has an early steep decay and late re-brightening; GRB 070714B does not have a very steep decay phase but has flaring along the way; GRB 061006 has an early steep decay but no late re-brightening), but can be all understood within the framework of a magnetar central engine. We find that with standard parameter values, the magnetar remnant scenario can well interpret the multi-band data of all three bursts, including the extended emission and their late chromatic features for X-ray and optical data. 

\begin{table}
\begin{center}{\scriptsize
\caption{Model parameters to interpret the broadband data of GRB 050724, GRB 070714B and GRB 061006}
\begin{tabular}{ccccccc} \hline\hline
 \multicolumn{7}{c}{Magnetar and ejecta parameters}\\
  \hline
 &$B~({\rm G})$                    & $P_i~({\rm ms})$   &$R_s~({\rm cm})$ &$M_{\rm ej}~({\rm M_{\odot}})$                    &  $v_i/c$   & $\kappa~({\rm cm^{2}~g^{-1}})$     \\
GRB 050724 & $6\times10^{15}$     &  $5$         &$1.2\times10^{6}$ & $10^{-3}$     &  $0.2$         &$1$\\
GRB 070714B & $1\times10^{16}$     &  $2.5$         &$1.0\times10^{6}$ & $10^{-2}$     &  $0.2$         &$1$\\
GRB 061006 & $5\times10^{15}$     &  $2$         &$1.0\times10^{6}$ & $10^{-2}$     &  $0.2$         &$1$\\
   \hline
  \multicolumn{7}{c}{Jet and ambient medium parameters}\\
  \hline
  & &$E~({\rm erg})$     &$\Gamma_0$               &  $n~(\rm{cm^{-3}})$& $\theta~({\rm rad})$ \\
 GRB 050724& &$3.9\times 10^{50}$     &  $200$ &  $0.1$   & $0.2$       \\
 GRB 070714B& &$10^{52}$     &  $95$ &  $0.01$   & $0.2$       \\
 GRB 061006& &$1.6\times 10^{52}$     &  $200$ &  $0.1$   & $0.2$       \\
  \hline
  \multicolumn{7}{c}{Other parameters}\\
  \hline
&& $\epsilon_e$                    &  $\epsilon_B$&  $p$ &  $\xi$ & \\
 GRB 050724& &$0.025$     &  $0.001$&  $2.3$ &  $0.01$ &    \\
 GRB 070714B &&$0.06$     &  $0.0002$&  $2.6$ &  $0.1$ &      \\
 GRB 061006& &$0.015$     &  $0.00003$&  $2.1$  & $0.01$ &   \\
   \hline\hline
 \end{tabular}
 }
\end{center}
\end{table}

The fact that all three internal-plateau short GRBs with redshift measurement and late X-ray/optical observations have merger-nova signatures suggest that short GRBs with internal plateaus are indeed powered by a magnetar central engine. We therefore encourage intense late-time multi-color optical follow-up observations of short GRBs with extended emission/internal plateau to identify more magnetar-powered merger-novae in the future.

It is interesting to compare the properties of magnetar-powered merger-novae and the r-process powered merger-novae claimed in the literature. In Figure 2, we present the peak luminosities of all claimed cases, compared with the typical luminosities of novae, supernovae, and super-luminous supernovae. One can see that the three r-process powered merger-novae associated with GRB 050709, GRB 060614, and GRB 130603B indeed have peak luminosities about 1000 times of that of a typical nova \footnote{Note that the kilonova following GRB 130603B is much more robust (with an excess of larger than 100 times of the luminosity expected from the afterglow \citep{tanvir13,berger13}) than the others.}. The three magnetar-powered merger-novae claimed in this paper, on the other hand, are systematically brighter by more than one order of magnitude, so that the term ``kilo-nova'' cannot catch the properties of these events. The two populations are clearly separated from each other. More late-time follow-up observations of short GRBs are needed to quantify the fraction of NS-NS mergers with a magnetar merger product.

\begin{figure}
\centering
\includegraphics[width=0.45\textwidth]{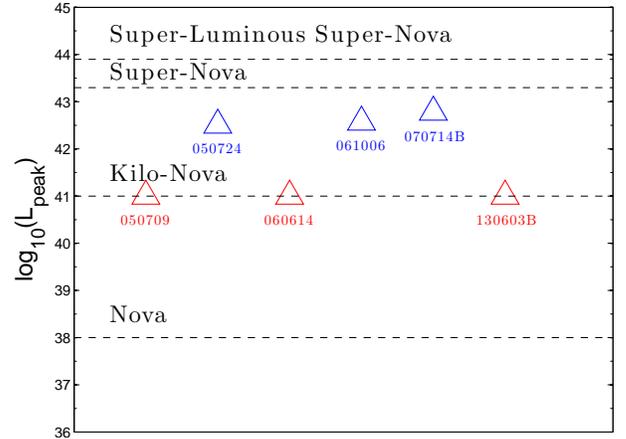}
\caption{Peak bolometric luminosity for all claimed ``kilo-novae" (GRB 050709 \citep{jin16}, GRB 060614 \citep{yang15}, GRB 130603B \citep{tanvir13,berger13}) and magnetar-powered merger-novae. It is clear that the three magnetar-powered merger-novae suggested in this paper are systematically brighter than the ``kilo-novae", by more than one order of magnitude.}
\label{fig:Lp}
\end{figure}

To interpret the data of GRB 050724, we adopt a relatively large value for the initial spin period of the magnetar $P_i=5~{\rm ms}$. It is worth noticing that this is just an effective way to remark that a relatively small energy budget for the magnetar power (e.g., $\sim 10^{51}~\rm{ergs}$) is required. In principle, the supra-massive or stable neutron star remnant should be rapidly spinning near the breakup limit (e.g., $P_i\sim1~{\rm ms}$ ), since before the merger the two NSs are in the Keplerian orbits. In this case, the total spin energy of the central star is of the order of $10^{52}$ erg. However, there are several channels to distribute the total spin energy, i.e., the 
EM radiation channel, the GW radiation channel, and the channel to fall into the BH for supramassive NSs. For GRB 050724 (also relevant to GRB 070714B and GRB 061006), the energy budget in the EM channel is smaller or even much smaller than $10^{52}$ erg, implying that a good fraction of the initial spin energy may be released in the form of GW radiation loss during \citep{radice16} or after the merger due to the large deformation of the magnetar \citep{fan13b,gao16,lasky16}, or falling into the BH. Most recently, \cite{fong16} studied the long-term radio behavior of GRB 050724 with the Very Large Array, and placed a stringent limit of $E_{\rm max}\approx(2-5)\times10^{51}~\rm erg$ on the rotational energy of a stable magnetar. This is consistent with our results. However, we notice that the limit in \cite{fong16} is placed by assuming $\epsilon_B=0.1$, a relatively extreme value for $\epsilon_B$ in GRB afterglow modeling \citep{kumarzhang15,wang15}. We adopt $\epsilon_B=0.001$ to interpret the afterglow data of GRB 050724, in this case, the constraint on the rotational energy would become much looser, e.g., $E_{\rm max}$ could be larger than $10^{52}~\rm erg$ \citep{fong16}.

Taking into account GRB 080503, we now have 4 candidates of magnetar-powered merger-nova. Among the sample, GRB 080503 and GRB 050724 show late re-brightening feature in the X-ray band, indicating a stable magnetar (at least stable up to $10^{5}$ s) as the central engine. For GRB 070714B and GRB 061006, the supra-massive NSs seem to have collapsed to black holes before their surrounding ejecta become transparent (collapse before $10^{5}$ s). Although the sample is small, the ratio between stable magnetars and supra-massive magnetars is roughly $1:1$, which is consistent with the observational results obtained by analyzing the X-ray behavior of particular short GRBs \citep{rowlinson13,gompertz14,fong16} and the theoretical results predicted by \cite{gao16}, where a neutron star EoS with a maximum mass close to a parameterization of $M_{\rm max} = 2.37\,M_\odot (1+1.58\times10^{-10} P^{-2.84})$ is adopted.  A larger sample of magnetar-powered merger-novae in the future could give more stringent constraints on the EoS for neutron matter.

With the current sample, some simple statistics may be obtained. For instance, for the central magnetar, the values of initial spin period spans from $2$ ms to $5$ ms, and the dipolar magnetic field of strength span from $5\times 10^{15}~{\rm G}$ to $10^{16}~{\rm G}$. The mass of the ejecta material spans from $10^{-3}~{\rm M_{\odot}}$  to $10^{-2}~{\rm M_{\odot}}$.  A larger sample in the future would increase the statistics and shed light into the detailed properties of the binary NS merger products, both the central magnetar and the surrounding ejecta.

\acknowledgments

We thank the referee for the helpful comments which have helped us to improve the presentation
of the paper. This work is supported by the National Basic Research Program (973 Program) of China (Grant No. 2014CB845800), and the National Natural Science Foundation of China under Grant No. 11543005, 11603003, 11603006,11633001,11690024. L. H. J. acknowledges support by the Scientific Research Foundation of Guangxi University (Grant No. XGZ150299) and One-Hundred-Talents Program of Guangxi colleges.

\appendix

Considering that one NS-NS merger event leaves behind a millisecond magnetar with an initial spin period $P_i$ and a dipolar magnetic field of strength $B$, surrounded by a quasi-spherical ejecta shell with mass $M_{\rm ej}$ and initial speed $v_i$. The total rotational energy of the magnetar reads $E_{\rm{rot}}=(1/2)I
\Omega_{0}^{2} \simeq 2\times 10^{52} I_{45} P_{i,-3}^{-2} ~{\rm
erg}$ (with $I_{45} \sim 1.5$ for a massive neutron star). The spin-down luminosity as a function of time could be expressed as 
\begin{eqnarray}
L_{\rm sd}=L_{\rm sd,i}\left(1+{t\over t_{\rm
sd}}\right)^{-2}
\end{eqnarray}
where
\begin{eqnarray}
L_{\rm sd,i}=10^{47}~R_{s,6}^6B_{14}^{2}P_{i,-3}^{-4}\rm~erg~s^{-1}
\end{eqnarray}
is the initial spin-down luminosity, and
\begin{eqnarray}
t_{\rm sd}=2\times10^{5}~R_{s,6}^{-6}B_{14}^{-2}P_{i,-3}^{2}~\rm s
\end{eqnarray}
is the spin-down timescale. In our treatment, we do not introduce the spindown term due to gravitational wave radiation, but allow $P_i$ be a free parameter, which can be longer than 1 ms. In terms of merger-nova dynamics, the result would be similar if one takes $P_i \sim 1$ ms but explicitly introduces a GW spindown term \citep{gao16,sun16}. 

In the directions not blocked by the ejecta (either intrinsic or drilled by the jet), internal dissipation of the magnetar wind 
would give rise to extended emission (along the jet direction) or a GRB-less X-ray emission in the off-axis direction \citep{zhang13}. Assuming an efficiency factor $\eta_{\nu}$ to convert the spin-down luminosity to the observed luminosity at frequency $\nu$, we have
\begin{eqnarray}
F_{\nu} & \sim & \frac{\eta_{\nu} L_{\rm sd}}{4 \pi \nu D_L^2}.
\end{eqnarray}
In other directions, the magnetar wind runs into the ejecta and is quickly decelerated. The dissipation photons (either from forced reconnection or self-dissipation) would be initially trapped and eventually show up when the ejecta becomes optically thin. On the other hand, the injected wind continuously pushes from behind and accelerates the ejecta, where the dynamical evolution of the ejecta can be determined by \citep{yu13}
\begin{eqnarray}
{d\Gamma\over dt}={{dE\over dt}-\Gamma {\cal D}\left({dE'_{\rm int}\over
dt'}\right)-(\Gamma^2-1)c^2\left({dM_{\rm sw}\over dt}\right)\over
M_{\rm ej}c^2+E'_{\rm int}+2\Gamma M_{\rm sw}c^2}
\label{eq:Gt}
\end{eqnarray}
where $\Gamma$ is the bulk Lorentz factor of the ejecta, $M_{\rm sw}=\frac{4\pi}{3}R^3nm_p$ is the
swept mass from the interstellar medium (with density $n$) and $R$ is the radius of the ejecta. ${\cal D}=1/[\Gamma(1-\beta)]$ is the Doppler factor with $\beta=\sqrt{1-\Gamma^{-2}}$. $E'_{\rm int}$ is the internal energy measured in the comoving rest frame, whose variation could be expressed as \cite[e.g.][]{kasen10,yu13}
\begin{eqnarray}
{dE'_{\rm int}\over dt'}=\xi {\cal D}^{-2}L_{\rm sd}+ L'_{\rm ra} -L'_{\rm e}
-\mathcal P'{dV'\over dt'},
\label{eq:Ep}
\end{eqnarray}
where $L'_{\rm ra}$ is the radioactive power, which could be estimated by
\begin{eqnarray}
L'_{\rm ra}=4\times10^{49}M_{\rm ej,-2}\left[{1\over2}-{1\over\pi}\arctan \left({t'-t'_0\over
t'_\sigma}\right)\right]^{1.3}~\rm erg~s^{-1},
\label{eq:Lrap}
\end{eqnarray}
with $t'_0 \sim 1.3$ s and $t'_\sigma \sim 0.11$ s \citep{korobkin12}. $L'_e$ is the radiated bolometric
luminosity, reading as \footnote{The energy loss due to shock emission is ignored here, as is usually done
in GRB afterglow modeling.}
\begin{eqnarray}
L'_e=\left\{
\begin{array}{l l}
  {E'_{\rm int}c\over \tau R/\Gamma}, & \tau>1, \\
  {E'_{\rm int}c\over R/\Gamma}, &\tau<1,\\ \end{array} \right.\
  \label{eq:Lep}
\end{eqnarray}
where $\tau=\kappa (M_{\rm ej}/V')(R/\Gamma)$ is the optical depth of the ejecta with $\kappa$ being
the opacity \citep{kasen10,kotera13}. $\mathcal P'=E'_{\rm int}/3V'$ is the radiation dominated pressure where the
comoving volume evolution can be fully addressed by
\begin{eqnarray}
{dV'\over dt'}=4\pi R^2\beta c,
\label{eq:Vp}
\end{eqnarray}
together with
\begin{eqnarray}
{dR\over dt}={\beta c\over (1-\beta)}.
\label{eq:rt}
\end{eqnarray}
 
With energy conservation, we have
\begin{eqnarray}
{dE\over dt}=\xi L_{\rm sd}+{\cal D}^{2}L'_{\rm ra}-{\cal D}^{2}L'_{\rm e}.
\label{eq:Et}
\end{eqnarray}

The dynamical description of the ejecta could be easily obtained by solving above differential equations. Assuming a blackbody spectrum for the thermal emission of the merger-nova, the observed flux for a given frequency $\nu$ could be calculated as
\begin{eqnarray}
F_{\nu}={1\over4\pi D_L^2\max(\tau,1)}{8\pi^2  {\cal D}^2R^2\over
h^3c^2\nu}{(h\nu/{\cal D})^4\over \exp(h\nu/{\cal D}kT')-1},
\end{eqnarray}
where $h$ is the Planck constant.

The interaction between the ejecta and the ambient medium would drive a strong external shock, where particles would be accelerated and broad-band synchrotron radiation emitted.  Assuming that the accelerated electrons is a power law function with the index of p (
$\frac{dN_{\rm e}'}{d\gamma_{\rm e}} \propto \gamma_{\rm e}^{-p},~\gamma_{\rm e,m}\leq \gamma_{\rm e} \leq\gamma_{\rm e,M}$), and that a constant fraction $\epsilon_e$ of the shock energy is distributed to electrons, the minimum injected electron Lorentz factor could be estimated as  ($p\geq 2$)
\begin{eqnarray}
\gamma_{\rm e,m}=g(p)\epsilon_e(\Gamma -1) \frac {m_p} {m_e},
\end{eqnarray}
where the function $g(p)$ takes the form
\begin{eqnarray}
\label{gp} g(p) \simeq \left\{ \begin{array}{ll} \frac{p-2}{p-1}, & p>2;\\
\rm{ln}^{-1}(\gamma_{\rm e,M}/\gamma_{\rm e,m}), &
p=2. \\
\end{array} \right.
\end{eqnarray}
The maximum electron Lorentz factor $\gamma_{e,M}$ could be estimated by balancing the acceleration time scale and the dynamical time scale, i.e.
\begin{eqnarray}
\gamma_{\rm e,M}\sim \frac{\Gamma t q_e B}{\zeta m_p c},
\end{eqnarray}
where $\zeta\sim 1$ is a parameter that describes the details of acceleration. Assuming that the magnetic energy density behind
the shock is a constant fraction $\epsilon_B$ of the shock energy density, one can obtain $B'=(8 \pi e_s\epsilon_B)^{1/2}$, where $B'$ is the comoving magnetic
field strength and $e_s$ is the energy density in the shocked region. 

According to standard synchrotron radiation model, in the co-moving frame, the synchrotron radiation power at frequency $\nu '$ from electrons is given by
(Rybicki \& Lightman 1979)
\begin{equation}
\label{eq:pnup}
P'_{\nu'} = \frac{\sqrt{3} q_e^3 B'}{m_{\rm e} c^2}
	    \int_{\gamma_{\rm e,m}}^{\gamma_{\rm e,M}}
	    \left( \frac{dN_{\rm e}'}{d\gamma_{\rm e}} \right)
	    F\left(\frac{\nu '}{\nu_{\rm cr}'} \right) d\gamma_{\rm e},
\end{equation}
where $q_e$ is electron charge,
$\nu_{\rm cr}' = 3 \gamma_{\rm e}^2 q_e B' / (4 \pi m_{\rm e} c)$ is the
characteristic frequency of an electron with Lorentz factor $\gamma_e$, and
\begin{equation}
F(x) = x \int_{x}^{+ \infty} K_{5/3}(k) dk,
\end{equation}
with $K_{5/3}(k)$ being the Bessel function. The observed flux density at frequency $\nu={\cal D}\nu'$ can be calculated as
\begin{eqnarray}
F_{\nu}={{\cal D}^3\over4\pi D_L^2}P'_{\nu'},
\label{eq:fnuobs}
\end{eqnarray}

The interaction between the GRB jet and the ambient medium could also generate a strong
external shock, giving rise to broad-band GRB afterglow emission \cite[][for a review]{gao13review}. The total effective kinetic energy of the jet and the medium can be expressed as
\begin{eqnarray}
E=(\Gamma-1)M_{\rm jet}c^2+(\Gamma^2-1)M_{\rm sw}c^2,
\end{eqnarray}
where $M_{\rm sw}=2\pi(1-\cos\theta)/3R^3nm_p$ with $\theta$ being the half opening angle of the
jet. The energy conservation law gives
\begin{eqnarray}
{d\Gamma\over dt}={-(\Gamma^2-1)\left({dM_{\rm sw}\over dt}\right)\over
M_{\rm jet}+2\Gamma M_{\rm sw}}.
\label{eq:dyngrb}
\end{eqnarray}
where the energy loss due to shock emission is ignored. With a dynamical solution for the jet and the aforementioned radiation equations, one can calculate the relevant broadband GRB afterglow emission.


\begin{thebibliography}{}
\expandafter\ifx\csname natexlab\endcsname\relax\def\natexlab#1{#1}\fi

\bibitem[{{Abbott} {et~al.}(2009){Abbott}, {Abbott}, {Adhikari}, {Ajith},
  {Allen}, {Allen}, {Amin}, {Anderson}, {Anderson}, {Arain}, \& et~al.}]{ligo}
{Abbott}, B.~P., {Abbott}, R., {Adhikari}, R., {et~al.} 2009, Reports on
  Progress in Physics, 72, 076901

\bibitem[Abbott et al.(2016b)]{abbott16b} Abbott, B.~P., Abbott, R., Abbott, T.~D., et al.\ 2016b, Physical Review Letters, 116, 241103 

\bibitem[Abbott et al.(2016a)]{abbott16a} Abbott, B.~P., Abbott, R., Abbott, T.~D., et al.\ 2016a, Physical Review Letters, 116, 061102 

\bibitem[{{Acernese} {et~al.}(2008){Acernese}, {Alshourbagy}, {Amico},
  {Antonucci}, {Aoudia}, {Astone}, {Avino}, {Baggio}, {Ballardin}, \&
  {Barone}}]{virgo}
{Acernese}, F., {Alshourbagy}, M., {Amico}, P., {et~al.} 2008, Classical and
  Quantum Gravity, 25, 114045
  
\bibitem[Barbier et al.(2007)]{GCN.6623} Barbier, L., Barthelmy, S.~D., Cummings, J., et al.\ 2007, GRB Coordinates Network, 6623, 1 

\bibitem[Barnes \& Kasen(2013)]{barnes13} Barnes, J., \& Kasen, D.\ 2013, \apj, 775, 18 

\bibitem[Barthelmy et al.(2005)]{barthelmy05} Barthelmy, S.~D., Chincarini, G., Burrows, D.~N., et al.\ 2005, \nat, 438, 994 

\bibitem[{{Bauswein} {et~al.}(2013){Bauswein}, {Goriely}, \&
  {Janka}}]{bauswein13}
{Bauswein}, A., {Goriely}, S., \& {Janka}, H.-T. 2013, \apj, 773, 78

\bibitem[Beniamini et al.(2015)]{beniamini15} Beniamini, P., Nava, L., Duran, R.~B., \& Piran, T.\ 2015, \mnras, 454, 1073 

\bibitem[Berger et al.(2005)]{berger05} Berger, E., Price, P.~A., Cenko, S.~B., et al.\ 2005, \nat, 438, 988 

\bibitem[Berger et al.(2007)]{berger07} Berger, E., Fox, D.~B., Price, P.~A., et al.\ 2007, \apj, 664, 1000 

\bibitem[Berger(2011)]{berger11} Berger, E.\ 2011, \nar, 55, 1 

\bibitem[{{Berger} {et~al.}(2013){Berger}, {Fong}, \& {Chornock}}]{berger13}
{Berger}, E., {Fong}, W., \& {Chornock}, R. 2013, \apjl, 774, L23

\bibitem[Bucciantini et al.(2012)]{bucciantini12} Bucciantini, N., Metzger, B.~D., Thompson, T.~A., \& Quataert, E.\ 2012, \mnras, 419, 1537 

\bibitem[Burenin et al.(2005)]{GCN.3671} Burenin, R., Pavlinsky, M., Sunyaev, R., et al.\ 2005, GRB Coordinates Network, 3671, 1

\bibitem[{{Burrows} {et~al.}(2006){Burrows}, {Grupe}, {Capalbi}, {Panaitescu},
  {Patel}, {Kouveliotou}, {Zhang}, {M{\'e}sz{\'a}ros}, {Chincarini}, {Gehrels},
  \& {Wijers}}]{burrows06}
{Burrows}, D.~N., {Grupe}, D., {Capalbi}, M., {et~al.} 2006, \apj, 653, 468


\bibitem[Campana et al.(2006)]{campana06} Campana, S., Tagliaferri, G., Lazzati, D., et al.\ 2006, \aap, 454, 113 

\bibitem[Cenko et al.(2008)]{cenko08} Cenko, S.~B., Berger, E., Nakar, E., et al.\ 2008, arXiv:0802.0874 

\bibitem[Chu et al.(2016)]{chu16} Chu, Q., Howell, E.~J., Rowlinson, A., et al.\ 2016, \mnras, 459, 121 

\bibitem[Connaughton et al.(2016)]{connaughton16} Connaughton, V., Burns, E., Goldstein, A., et al.\ 2016, arXiv:1602.03920 

\bibitem[Covino et al.(2005)]{GCN.3665} Covino, S., Antonelli, L.~A., Romano, P., et al.\ 2005, GRB Coordinates Network, 3665, 1 

\bibitem[{{Dai} {et~al.}(2006){Dai}, {Wang}, {Wu}, \& {Zhang}}]{dai06}
{Dai}, Z.~G., {Wang}, X.~Y., {Wu}, X.~F., \& {Zhang}, B. 2006, Science, 311,
  1127

\bibitem[D'Avanzo et al.(2005)]{GCN.3690} D'Avanzo, P., Covino, S., Antonelli, L.~A., et al.\ 2005, GRB Coordinates Network, 3690, 1 

\bibitem[{{De Pasquale} {et~al.}(2010){De Pasquale}, {Schady}, {Kuin}, {Page},
  {Curran}, {Zane}, {Oates}, {Holland}, {Breeveld}, \&
  {Hoversten}}]{depasquale10}
{De Pasquale}, M., {Schady}, P., {Kuin}, N.~P.~M., {et~al.} 2010, \apjl, 709,
  L146

\bibitem[{{Dessart} {et~al.}(2009){Dessart}, {Ott}, {Burrows}, {Rosswog}, \&
  {Livne}}]{dessart09}
{Dessart}, L., {Ott}, C.~D., {Burrows}, A., {Rosswog}, S., \& {Livne}, E. 2009,
  \apj, 690, 1681

\bibitem[{{Eichler} {et~al.}(1989){Eichler}, {Livio}, {Piran}, \&
  {Schramm}}]{eichler89}
{Eichler}, D., {Livio}, M., {Piran}, T., \& {Schramm}, D.~N. 1989, \nat, 340,
  126

\bibitem[Evans et al.(2009)]{evans09} Evans, P.~A., Beardmore, A.~P., Page, K.~L., et al.\ 2009, \mnras, 397, 1177 

\bibitem[{{Fan} \& {Xu}(2006)}]{fanxu06}
{Fan}, Y.-Z., \& {Xu}, D. 2006, \mnras, 372, L19


\bibitem[{{Fan} {et~al.}(2013{\natexlab{a}}){Fan}, {Yu}, {Xu}, {Jin}, {Wu},
  {Wei}, \& {Zhang}}]{fan13a}
{Fan}, Y.-Z., {Yu}, Y.-W., {Xu}, D., {et~al.} 2013{\natexlab{a}}, \apjl, 779,
  L25
  
 \bibitem[{{Fan} {et~al.}(2013{\natexlab{b}}){Fan}, {Wu}, \& {Wei}}]{fan13b}
{Fan}, Y.-Z., {Wu}, X.-F., \& {Wei}, D.-M. 2013{\natexlab{b}}, \prd, 88, 067304

\bibitem[{{Fern{\'a}ndez} \& {Metzger}(2013)}]{fernandez13}
{Fern{\'a}ndez}, R., \& {Metzger}, B.~D. 2013, \mnras, 435, 502

\bibitem[Fong et al.(2016)]{fong16} Fong, W.-F., Metzger, B.~D., Berger, E., \& Ozel, F.\ 2016a, arXiv:1607.00416 

\bibitem[Fong et al.(2016b)]{fong16b} Fong, W.-F., Margutti, R., Chornock, R., et al.\ 2016b, arXiv:1608.08626 

\bibitem[Fryer et al.(2015)]{fryer15} Fryer, C.~L., Belczynski, K., Ramirez-Ruiz, E., et al.\ 2015, \apj, 812, 24 

\bibitem[Gal-Yam et al.(2006)]{gal-yam06} Gal-Yam, A., Fox, D.~B., Price, P.~A., et al.\ 2006, \nat, 444, 1053 

\bibitem[{{Gao} {et~al.}(2013a){Gao}, {Ding}, {Wu}, {Zhang}, \&
  {Dai}}]{gao13GWB}
{Gao}, H., {Ding}, X., {Wu}, X.-F., {Zhang}, B., \& {Dai}, Z.-G.
  2013a, \apj, 771, 86

\bibitem[Gao et al.(2013b)]{gao13review} Gao, H., Lei, W.-H., Zou, Y.-C., Wu, X.-F., \& Zhang, B.\ 2013b, \nar, 57, 141 

\bibitem[Gao et al.(2015)]{gao15} Gao, H., Ding, X., Wu, X.-F., Dai, Z.-G., \& Zhang, B.\ 2015, \apj, 807, 163 

\bibitem[Gao et al.(2016)]{gao16} Gao, H., Zhang, B., L{\"u}, H.-J.\ 2016, \prd, 93, 044065 

\bibitem[{{Gao} \& {Fan}(2006)}]{gaofan06}
{Gao}, W.-H., \& {Fan}, Y.-Z. 2006, Chinese Journal of Astronomy and
  Astrophysics, 6, 513

\bibitem[Gehrels et al.(2005)]{gehrels05} Gehrels, N., Sarazin, C.~L., O'Brien, P.~T., et al.\ 2005, \nat, 437, 851 

\bibitem[Gehrels et al.(2006)]{gehrels06} Gehrels, N., Norris, J.~P., Barthelmy, S.~D., et al.\ 2006, \nat, 444, 1044 

\bibitem[{{Giacomazzo} \& {Perna}(2013)}]{giacomazzo13}
{Giacomazzo}, B., \& {Perna}, R. 2013, \apjl, 771, L26

\bibitem[Gompertz et al.(2014)]{gompertz14} Gompertz,
    B.~P., O'Brien, P.~T., \& Wynn, G.~A.\ 2014, \mnras, 438,
    240


\bibitem[Gompertz et al.(2013)]{gompertz13} Gompertz,
    B.~P., O'Brien, P.~T., Wynn, G.~A., \& Rowlinson, A.\ 2013,
    \mnras, 431, 1745

\bibitem[Graham et al.(2007)]{GCN.6836} Graham, J.~F., Fruchter, A.~S., Levan, A.~J., et al.\ 2007, GRB Coordinates Network, 6836, 1

\bibitem[Graham et al.(2009)]{graham09} Graham, J.~F., Fruchter, A.~S., Levan, A.~J., et al.\ 2009, \apj, 698, 1620 
 
\bibitem[Greiner et al.(2016)]{greiner16} Greiner, J., Burgess, J.~M., Savchenko, V., \& Yu, H.-F.\ 2016, arXiv:1606.00314 

\bibitem[Horesh et al.(2016)]{horesh16} Horesh, A., Hotokezaka, K., Piran, T., Nakar, E., \& Hancock, P.\ 2016, \apjl, 819, L22 

\bibitem[{{Hotokezaka} {et~al.}(2013){Hotokezaka}, {Kiuchi}, {Kyutoku},
  {Okawa}, {Sekiguchi}, {Shibata}, \& {Taniguchi}}]{hotokezaka13}
{Hotokezaka}, K., {Kiuchi}, K., {Kyutoku}, K., {et~al.} 2013, \prd, 87, 024001


\bibitem[Hurley et al.(2006)]{GCN.5702} Hurley, K., Cline, T., Smith, D.~M., et al.\ 2006, GRB Coordinates Network, 5702, 1 

\bibitem[Jin et al.(2016)]{jin16} Jin, Z.-P., Hotokezaka, K., Li, X., et al.\ 2016, arXiv:1603.07869 

\bibitem[Kasen \& Bildsten(2010)]{kasen10} Kasen, D., \& Bildsten, L.\ 2010, \apj, 717, 245 

\bibitem[{{Kasen} {et~al.}(2013){Kasen}, {Badnell}, \& {Barnes}}]{kasen13}
{Kasen}, D., {Badnell}, N.~R., \& {Barnes}, J. 2013, \apj, 774, 25

\bibitem[Kasen et al.(2015)]{kasen15} Kasen, D., Fern{\'a}ndez, R., \& Metzger, B.~D.\ 2015, \mnras, 450, 1777 

\bibitem[Korobkin et al.(2012)]{korobkin12} Korobkin, O., Rosswog, S., Arcones, A., \& Winteler, C.\ 2012, \mnras, 426, 1940 

\bibitem[Kotera et al.(2013)]{kotera13} Kotera, K., Phinney, E.~S., \& Olinto, A.~V.\ 2013, \mnras, 432, 3228 

\bibitem[Krimm et al.(2005)]{GCN.3667} Krimm, H., Barbier, L., Barthelmy, S., et al.\ 2005, GRB Coordinates Network, 3667, 1 

\bibitem[{{Kulkarni}(2005)}]{kulkarni05}
{Kulkarni}, S.~R. 2005, ArXiv Astrophysics e-prints, astro-ph/0510256

\bibitem[Kumar \& Zhang(2015)]{kumarzhang15} Kumar, P., \& Zhang, B.\ 2015, \physrep, 561, 1
  
\bibitem[{{Kuroda} \& {LCGT Collaboration}(2010)}]{kagra}
{Kuroda}, K., \& {LCGT Collaboration}. 2010, Classical and Quantum Gravity, 27,
  084004

\bibitem[Lasky et al.(2014)]{lasky14} Lasky, P.~D., Haskell, B., Ravi, V., Howell, E.~J., \& Coward, D.~M.\ 2014, \prd, 89, 047302 

\bibitem[Lasky \& Glampedakis(2016)]{lasky16} Lasky, P.~D., \& Glampedakis, K.\ 2016, \mnras, 458, 1660 

\bibitem[{{Lee} {et~al.}(2009){Lee}, {Ramirez-Ruiz}, \&
  {L{\'o}pez-C{\'a}mara}}]{lee09}
{Lee}, W.~H., {Ramirez-Ruiz}, E., \& {L{\'o}pez-C{\'a}mara}, D. 2009, \apjl,
  699, L93
  
\bibitem[Levan et al.(2007)]{GCN.6630}  Levan, A. J., Tanvir, N. R., Bonfield, D., Martinez-Sansigre, A., Graham, J., \& Fruchter, A. 2007, GRB Coordinates Network, 6630, 1

\bibitem[{{Li} \& {Paczy{\'n}ski}(1998)}]{lipaczynski98}
{Li}, L.-X., \& {Paczy{\'n}ski}, B. 1998, \apjl, 507, L59

\bibitem[Li et al.(2016)]{li16} Li, A., Zhang, B., Zhang, N.-B., et al.\ 2016, \prd, 94, 083010 

\bibitem[Liang et al.(2007)]{liang07} Liang, E.-W., Zhang, B.-B., \& Zhang, B.\ 2007, \apj, 670, 565 

\bibitem[Loeb(2016)]{loeb16} Loeb, A.\ 2016, \apjl, 819, L21 

\bibitem[{{L{\"u}} \& {Zhang}(2014)}]{lv14}
{L{\"u}}, H.-J., \& {Zhang}, B. 2014, \apj, 785, 74

\bibitem[{{L{\"u}} {et~al.}(2015){L{\"u}}, {Zhang}, {Lei}, {Li}, \&
  {Lasky}}]{lv15}
{L{\"u}}, H.-J., {Zhang}, B., {Lei}, W.-H., {Li}, Y., \& {Lasky}, P.~D. 2015,
  \apj, 805, 89

\bibitem[Malesani et al.(2006a)]{GCN.5705} Malesani, D., Stella, L., Covino, S., et al.\ 2006a, GRB Coordinates Network, 5705, 1 

\bibitem[Malesani et al.(2006b)]{GCN.5718} Malesani, D., Stella, L., D'Avanzo, P., et al.\ 2006b, GRB Coordinates Network, 5718, 1 

\bibitem[Malesani et al.(2007)]{malesani07} Malesani, D., Covino, S., D'Avanzo, P., et al.\ 2007, \aap, 473, 77 

\bibitem[Margutti et al.(2011)]{margutti11} Margutti, R., Chincarini, G., Granot, J., et al.\ 2011, \mnras, 417, 2144

\bibitem[{{Metzger} {et~al.}(2008){Metzger}, {Quataert}, \&
  {Thompson}}]{metzger08}
{Metzger}, B.~D., {Quataert}, E., \& {Thompson}, T.~A. 2008, \mnras, 385, 1455

\bibitem[{{Metzger} {et~al.}(2010){Metzger}, {Mart{\'{\i}}nez-Pinedo},
  {Darbha}, {Quataert}, {Arcones}, {Kasen}, {Thomas}, {Nugent}, {Panov}, \&
  {Zinner}}]{metzger10}
{Metzger}, B.~D., {Mart{\'{\i}}nez-Pinedo}, G., {Darbha}, S., {et~al.} 2010,
  \mnras, 406, 2650

\bibitem[{{Metzger} \& {Berger}(2012)}]{metzger12}
{Metzger}, B.~D., \& {Berger}, E. 2012, \apj, 746, 48

\bibitem[{{Metzger} \& {Piro}(2014)}]{metzger14}
{Metzger}, B.~D., \& {Piro}, A.~L. 2014a, \mnras, 439, 3916

\bibitem[Metzger 
\& Fern{\'a}ndez(2014)]{metzger14b} Metzger, B.~D., \& Fern{\'a}ndez, R.\ 2014b, \mnras, 441, 3444

\bibitem[Metzger \& Bower(2014)]{metzger14c} Metzger, B.~D., \& Bower, G.~C.\ 2014c, \mnras, 437, 1821 

\bibitem[Nagakura et al.(2014)]{nagakura14} Nagakura, H., Hotokezaka, K., Sekiguchi, Y., Shibata, M., \& Ioka, K.\ 2014, \apjl, 784, L28 

\bibitem[{{Nakar} \& {Piran}(2011)}]{nakar11}
{Nakar}, E., \& {Piran}, T. 2011, \nat, 478, 82

\bibitem[Narayan et al.(1992)]{narayan92} Narayan, R., Paczynski, B., \& Piran, T.\ 1992, \apjl, 395, L83 

\bibitem[Norris \& Bonnell(2006)]{norris06} Norris,
    J.~P., \& Bonnell, J.~T.\ 2006, \apj, 643, 266

\bibitem[O'Brien et al.(2006)]{obrien06} O'Brien, P.~T., Willingale, R., Osborne, J., et al.\ 2006, \apj, 647, 1213 

\bibitem[Perley et al.(2007)]{GCN.6652}  Perley, D. A., Bloom, J. S., Thoene, C. , et al.\ 2007, GRB Coordinates Network, 6652, 1

\bibitem[Perley et al.(2009)]{perley09} Perley, D.~A., Metzger, B.~D., Granot, J., et al.\ 2009, \apj, 696, 1871 

\bibitem[Perna et al.(2016)]{perna16} Perna, R., Lazzati, D., \& Giacomazzo, B.\ 2016, \apjl, 821, L18 

\bibitem[{{Piran} {et~al.}(2013){Piran}, {Nakar}, \& {Rosswog}}]{piran13}
{Piran}, T., {Nakar}, E., \& {Rosswog}, S. 2013, \mnras, 430, 2121

\bibitem[Prochaska et al.(2005a)]{GCN.3679} Prochaska, J.~X., Chen, H.-W., Bloom, J.~S., \& Stephens, A.\ 2005a, GRB Coordinates Network, 3679, 1 

\bibitem[Prochaska et al.(2005b)]{GCN.3700} Prochaska, J.~X., Bloom, J.~S., Chen, H.-W., et al.\ 2005b, GRB Coordinates Network, 3700, 1 

\bibitem[Racusin et al.(2007a)]{GCN.6620} Racusin, J.~L., Barthelmy, S.~D., Burrows, D.~N., et al.\ 2007a, GRB Coordinates Network, 6620, 1 

\bibitem[{{Racusin} {et~al.}(2007{\natexlab{b}}){Racusin}, {Kennea}, {Pagani},
  {Vetere}, \& {Evans}}]{GCN.6627}
{Racusin}, J., {Kennea}, J., {Pagani}, C., {Vetere}, L., and {Evans}, P.
  2007{\natexlab{b}}, {GCN Circular} 6627
   
\bibitem[Radice et al.(2016)]{radice16} Radice, D., Bernuzzi, S., \& Ott, C.~D.\ 2016, arXiv:1603.05726 
  
\bibitem[Rezzolla et al.(2010)]{rezzolla10} Rezzolla, L., Baiotti, L., Giacomazzo, B., Link, D., \& Font, J.~A.\ 2010, Classical and Quantum Gravity, 27, 114105 

\bibitem[{{Rezzolla} {et~al.}(2011){Rezzolla}, {Giacomazzo}, {Baiotti},
  {Granot}, {Kouveliotou}, \& {Aloy}}]{rezzolla11}
{Rezzolla}, L., {Giacomazzo}, B., {Baiotti}, L., {et~al.} 2011, \apjl, 732, L6

\bibitem[Rosswog et al.(2000)]{rosswog00} Rosswog, S., Davies, M.~B., Thielemann, F.-K., \& Piran, T.\ 2000, \aap, 360, 171 

\bibitem[{{Rosswog} {et~al.}(2013){Rosswog}, {Piran}, \& {Nakar}}]{rosswog13}
{Rosswog}, S., {Piran}, T., \& {Nakar}, E. 2013, \mnras, 430, 2585

\bibitem[Rosswog et al.(2014)]{rosswog14} Rosswog, S., Korobkin, O., Arcones, A., Thielemann, F.-K., \& Piran, T.\ 2014, \mnras, 439, 744 

\bibitem[{{Rowlinson} {et~al.}(2013){Rowlinson}, {O'Brien}, {Metzger},
  {Tanvir}, \& {Levan}}]{rowlinson13}
{Rowlinson}, A., {O'Brien}, P.~T., {Metzger}, B.~D., {Tanvir}, N.~R., \&
  {Levan}, A.~J. 2013, \mnras, 430, 1061

\bibitem[{{Rowlinson} {et~al.}(2010){Rowlinson}, {O'Brien}, {Tanvir}, {Zhang},
  {Evans}, {Lyons}, {Levan}, {Willingale}, {Page}, {Onal}, {Burrows},
  {Beardmore}, {Ukwatta}, {Berger}, {Hjorth}, {Fruchter}, {Tunnicliffe}, {Fox},
  \& {Cucchiara}}]{rowlinson10}
{Rowlinson}, A., {O'Brien}, P.~T., {Tanvir}, N.~R., {et~al.} 2010, \mnras, 409,
  531
  
\bibitem[Rybicki \& Lightman(1979)]{1979rpa..book.....R} Rybicki, G.~B., \& Lightman, A.~P.\ 1979, New York, Wiley-Interscience, 1979.~393 p.,  


\bibitem[Santana et al.(2014)]{santana14} Santana, R., Barniol Duran, R., \& Kumar, P.\ 2014, \apj, 785, 29 

\bibitem[Sakamoto et al.(2011)]{sakamoto11} Sakamoto,
    T., Barthelmy, S.~D., Baumgartner, W.~H., et al.\ 2011,
    \apjs, 195, 2

\bibitem[Schady et al.(2006)]{schady06} Schady, P., Breeveld, A., Poole, T.~S., et al.\ 2006, GCN Report, 6, 1 

\bibitem[Siegel \& Ciolfi(2016a)]{siegel16a} Siegel, D.~M., \& Ciolfi, R.\ 2016a, \apj, 819, 14 

\bibitem[Siegel \& Ciolfi(2016b)]{siegel16b} Siegel, D.~M., \& Ciolfi, R.\ 2016b, \apj, 819, 15 

\bibitem[Sun et al.(2016)]{sun16} Sun, H., Zhang, B., \& Gao, H.\ 2016, arXiv:1610.03860 

\bibitem[{{Tanaka} \& {Hotokezaka}(2013)}]{tanaka13}
{Tanaka}, M., \& {Hotokezaka}, K. 2013, \apj, 775, 113

\bibitem[{{Tanvir} {et~al.}(2013){Tanvir}, {Levan}, {Fruchter}, {Hjorth},
  {Hounsell}, {Wiersema}, \& {Tunnicliffe}}]{tanvir13}
{Tanvir}, N.~R., {Levan}, A.~J., {Fruchter}, A.~S., {et~al.} 2013, \nat, 500,
  547

\bibitem[{{Wang} \& {Dai}(2013)}]{wanglj13}
{Wang}, L.-J., \& {Dai}, Z.-G. 2013, \apjl, 774, L33

\bibitem[Wang et al.(2015)]{wang15} Wang, X.-G., Zhang, B., Liang, E.-W., et al.\ 2015, \apjs, 219, 9 


\bibitem[Willingale et al.(2007)]{willingale07} Willingale, R., O'Brien, P.~T., Osborne, J.~P., et al.\ 2007, \apj, 662, 1093 

\bibitem[{{Wu} {et~al.}(2014){Wu}, {Gao}, {Ding}, {Zhang}, {Dai}, \&
  {Wei}}]{wu14}
{Wu}, X.-F., {Gao}, H., {Ding}, X., {et~al.} 2014, \apjl, 781, L10

\bibitem[Xiong(2016)]{xiong16} Xiong, S.\ 2016, arXiv:1605.05447 

\bibitem[Yang et al.(2015)]{yang15} Yang, B., Jin, Z.-P., Li,
X., et al.\ 2015, arXiv:1503.07761

\bibitem[Yu et al.(2013)]{yu13} Yu, Y.-W., Zhang, B., 
\& Gao, H.\ 2013, \apjl, 776, L40 

\bibitem[Yu et al.(2015)]{yu15} Yu, Y.-W., Li, S.-Z., \& Dai, Z.-G.\ 2015, \apjl, 806, L6 

\bibitem[{{Zhang} {et~al.}(2007){Zhang}, {Liang}, \& {Zhang}}]{zhangbb07}
{Zhang}, B.-B., {Liang}, E.-W., \& {Zhang}, B. 2007, \apj, 666, 1002

\bibitem[{{Zhang} {et~al.}(2009){Zhang}, {Zhang}, {Liang}, \&
  {Wang}}]{zhangbb09}
{Zhang}, B.-B., {Zhang}, B., {Liang}, E.-W., \& {Wang}, X.-Y. 2009, \apjl, 690,
  L10
  
\bibitem[Zhang et al.(2007)]{zhang07} Zhang, B., Zhang, B.-B., Liang, E.-W., et al.\ 2007, \apjl, 655, L25 


\bibitem[{{Zhang} \& {Yan}(2011)}]{zhangyan11}
{Zhang}, B., \& {Yan}, H. 2011, \apj, 726, 90

\bibitem[{{Zhang}(2013)}]{zhang13}
{Zhang}, B. 2013, \apjl, 763, L22

\bibitem[Zhang(2016)]{zhang16} Zhang, B.\ 2016, \apjl, 827, L31 

\bibitem[{Zhang \& Dai (2010)}]{zhangd10} Zhang, D., \& Dai, Z. G. 2010, \apj, 718, 841

\bibitem[Zhang et al.(2016)]{zhangsn16} Zhang, S.-N., Liu, Y., Yi, S., Dai, Z., \& Huang, C.\ 2016, arXiv:1604.02537 

\end{thebibliography}
\end{document}